\documentclass[12pt]{iopart}
\usepackage{iopams}
\usepackage{longtable}
\usepackage{amssymb}
\usepackage{graphicx}
\usepackage{dcolumn}

\begin{document}

\title{Analytical solutions for the Bohr Hamiltonian with the 
Woods--Saxon potential}

\author{M \c{C}apak$^1$, D Petrellis$^2$, B G\"{o}n\"{u}l$^1$ and Dennis Bonatsos$^3$}

\address{$^1$ Department of Engineering Physics, University of Gaziantep, 27310 Gaziantep, 
Turkey}

\address{$^2$ Department of Physics, University of Istanbul, 34134 Vezneciler, Istanbul, 
Turkey}

\address{$^3$ Institute of Nuclear and Particle Physics, National Centre for Scientific Research 
``Demokritos'', GR-15310 Aghia Paraskevi, Attiki, Greece}

\ead{gonul@gantep.edu.tr}

\begin{abstract}

Approximate analytical solutions in closed form are obtained for the 5-dimensional Bohr Hamiltonian with the Woods--Saxon potential, taking advantage of the Pekeris approximation and the exactly soluble one-dimensional extended Woods--Saxon 
potential with a dip near its surface. Comparison to the data for several $\gamma$-unstable and prolate deformed nuclei 
indicates that the potential can describe well the ground state and $\gamma_1$ bands of many prolate deformed nuclei corresponding to large enough ``well size'' and diffuseness, while it fails in describing the $\beta_1$ bands, due to its lack of a hard core, as well as in describing $\gamma$-unstable nuclei, because of the small ``well size'' and diffuseness they exhibit. 

\end{abstract}


\maketitle

\section{Introduction}

The advent of critical point symmetries \cite{IacE5,IacX5}, related to shape/phase transitions 
in nuclear structure \cite{Jolie,McCutchan}, has stirred interest in analytical solutions (exact or approximate) \cite{Fortunato}
of the Bohr Hamiltonian \cite{Bohr}. 
In addition to the infinite well potential, used in the critical point symmetries E(5) \cite{IacE5} and X(5) \cite{IacX5}, 
related to the transition from spherical to $\gamma$-unstable (soft with respect to triaxiality) \cite{Wilets} 
and to prolate deformed \cite{BM} nuclei respectively,  
solutions involving the Davidson \cite{Dav,Rohoz,EEP,ESD,DDMD}, Kratzer \cite{Kratzer,FV1,FV2,DDMK}, and Morse \cite{Morse,Boztosun,Inci} potentials (shown in figure 1) have been given.
When applied to the bulk of nuclei for which energy spectra and B(E2) transition rates are experimentally known \cite{ENSDF},
all these potentials provide good and quite similar results \cite{ESD,DDMD,FV1,FV2,DDMK,Boztosun,Inci}, despite having quite different shapes. 

The question arises if the success of the above mentioned potentials is due to the form of the Bohr Hamiltonian alone,
or if there are potentials which, when plugged into the Bohr Hamiltonian, will not be able to provide satisfactory 
results for nuclear data. 

In this work we consider the Woods--Saxon potential \cite{WS}, which has been extensively used in nuclear physics 
in a  different context, namely as a single-particle potential \cite{Heyde}. The potential, shown in figure 1(c), reads 
\begin{equation}\label{WSpot}
U_{WS}(\beta) = -{U_0\over e^{2a(\beta-\beta_0)} +1},
\end{equation}
where $U_0$, $a$ and $\beta_0$ are non-negative free parameters. 
This study is motivated by several reasons.

1) For $a\to \infty$ the WS potential reduces to a finite square well potential,
which has also been used in relation to critical point symmetries \cite{finite,Gonen}, 
in addition to the infinite well potential \cite{IacE5,IacX5}. 

2) In contrast to the Davidson \cite{Dav}, Kratzer \cite{Kratzer}, and Morse \cite{Morse} potentials, 
which possess a hard core, the WS potential does not have a hard core. We will find out the consequences of this difference.
 
3) The Woods--Saxon potential is also interesting from the mathematical point of view, since it 
has no closed analytical solution for the spectrum, even in one dimension and for vanishing angular momentum \cite{Koksal,Flugge}. Only an analytical wavefunction is known in one dimension and for vanishing angular 
momentum \cite{Flugge}.

\begin{figure*}[hbt]
\includegraphics[width=1\textwidth]{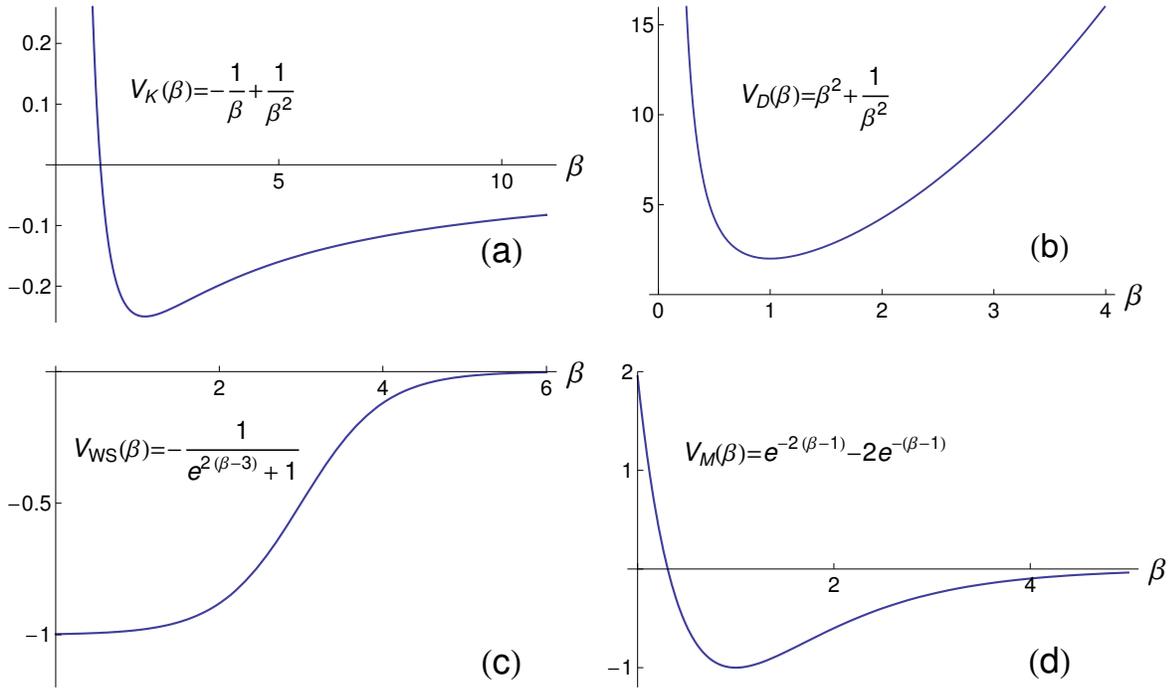}
\caption{(Color online) The Kratzer \cite{Kratzer} (a), Davidson \cite{Dav} (b), Woods--Saxon \cite{WS} (c)
and Morse \cite{Morse} (d) potentials, for special values of their free parameters, used for simplicity.
All quantities shown are dimensionless.  }
\end{figure*}
 
First we manage to produce approximate solutions in closed form for the WS potential with a centrifugal barrier,
exploiting its resemblance (after applying the Pekeris approximation \cite{Pekeris} to it) to a modified spherically symmetric 
WS potential \cite{Koura} known to possess exact analytical solutions \cite{Koksal}. 
Taking advantage of these solutions within the framework
of the Bohr Hamiltonian we subsequently fit several $\gamma$-unstable and prolate deformed nuclei, 
pointing out the successes, but also the failures of the WS potential. 

Numerical calculations involving generalized forms of the Bohr Hamiltonian have a long history. 

1) Extensive early numerical calculations for vibrating axially symmetric deformed nuclei 
have been performed in the framework of the Rotation Vibration Model \cite{Faessler,Sheline}.

2) Numerical solutions have subsequently been provided using a general form for the potential energy 
\cite{Kumar,Baranger}, giving emphasis on the restrictions imposed on the form of the potentials 
by symmetry constraints. 

3) Extensive numerical results using a general collective model employing general forms 
of both the kinetic energy and the potential energy have been obtained initially \cite{Gneuss}
by using a basis provided by Hecht \cite{Hecht}, and subsequently \cite{Hess1,Hess2}
by exploiting the basis of a 5-dimensional (5D) harmonic oscillator \cite{Chacon1,Chacon2}. 

4) The recent clarification \cite{Bahri} of the group theoretical structure of the Davidson potential, when used 
in the framework of the Bohr Hamiltonian, led to the development of the Algebraic Collective Model 
\cite{Rowe735,Rowe753,Caprio,Welsh}, which allows the efficient numerical calculation of spectra and transition probabilities 
of nuclei of any shape. 

In addition to the $\gamma$-unstable and the prolate deformed nuclei, 
the  description of triaxial nuclei within the framework of the Bohr Hamiltonian was very early attempted 
\cite{Filippov,Rostovsky} and is still attracting 
considerable attention \cite{Gonen,Yigit,Buganu,Inci23,Buganu91}.

On the other hand, studies on the microscopic foundation of the Bohr Hamiltonian have a long history, 
some early efforts reported in \cite{KB122,Kaniowska,Srebrny}. Efforts towards the derivation of the Bohr collective Hamiltonian through the adiabatic approximation of the time-dependent Hartree-Fock-Bogolyubov theory (the ATDHFB method),
as well as through the generator coordinate method with the Gaussian overlap appoximation (the GCM+GOA method) have been 
reviewed in \cite{Prochniak}. Recently, the derivation of the parameters to be used in a 5D Hamiltonian for quadrupole 
vibrational and rotational degrees of freedom from Relativistic Mean Field (RMF) calculations  \cite{ZPLi1,ZPLi2,ZPLi3,ZPLi4}
is acquiring momentum.


\section{Modified Woods--Saxon potential with a centrifugal barrier} 

\subsection{Pekeris approximation for the centrifugal barrier} 

The Pekeris approximation \cite{Pekeris} is a well known method for finding approximate analytical solutions 
for potentials involving exponentials. It has been introduced in relation to the Morse potential \cite{Morse} 
with a centrifugal barrier in studies of rotational-vibrational spectra of diatomic molecules \cite{Pekeris}. 
The basic idea of the Pekeris approximation is to rewrite approximately the centrifugal term using 
exponentials resembling the ones appearing in the rest of the potential, with the aim to be able to ``absorb'' 
the centrifugal term into the potential. 

This approximation has been used in relation with the Morse potential in diatomic molecules 
\cite{Pekeris,Berkdemir}, as well as with the pseudo-centrifugal term for Dirac particles within Morse potentials \cite{Bayrak}.
Furthermore, a supersymmetric improvement of the Pekeris approximation in the case of the Morse potential 
has been worked out \cite{Morales}. Numerical results reported in these papers indicate that the approximation
works well, especially for relatively low-lying vibrational and rotational states (see in particular the detailed 
results reported in \cite{Morales}). 

The Pekeris approximation has further been used in relation to various exponential-type potentials \cite{Dong},
to the Wei Hua oscillator \cite{Bera}, to the Rosen--Morse and Manning--Rosen potentials 
\cite{Ferreira}, as well as with the pseudo-centrifugal term for Dirac particles within Rosen--Morse potentials \cite{Wei}. 
In these cases, various functions appear in the original potentials, as well as in the Pekeris approximation terms. 

The spherically symmetric Woods-Saxon potential in three-dimensions with a centrifugal barrier reads 
\begin{equation}\label{WSbarrier}
U(\beta) = U_{WS}(\beta) + U_c(\beta) = -{U_0\over e^{2a(\beta-\beta_0)} +1} + {l(l+1)\over \beta^2},
\end{equation}
where $U_{WS}$ stands for the Woods-Saxon potential, $U_c$ stands for the centrifugal term, 
$l$ is the angular momentum, while $U_0$, $a$ and $\beta_0$ are non-negative free parameters. 

It is known \cite{Koksal} that this potential cannot be solved exactly. However, closed analytical solutions 
can be derived using the Pekeris approximation \cite{Pekeris}. As it was mentioned above, the basic idea of the Pekeris approximation is to rewrite approximately the centrifugal term using exponentials resembling the ones appearing in the rest of the potential, with the aim to be able to ``absorb'' the centrifugal term into the potential. 

In the present case we wish to write the centrifugal term 
in the approximate form  
\begin{equation}\label{Peker}
U_c(\beta)= {l(l+1)\over \beta^2} \approx \delta\left[ C_0 + {C_1\over e^{2a(\beta-\beta_0)}+1} +  
{C_2\over [e^{2a(\beta-\beta_0)}+1]^2} \right].
\end{equation}

In order to use a more compact notation, we introduce the change of coordinates 
$x= (\beta-\beta_0)/\beta_0$. Then the exact centrifugal term takes the form 
\begin{equation}
U_c(x) = {\delta \over (1+x)^2},
\end{equation}
where
\begin{equation}\label{delt}
\delta={l(l+1)\over \beta_0^2}. 
\end{equation}
Expanding this function into binomial series one gets 
\begin{equation}\label{Taylor1}
U_c(x) = \delta (1- 2x + 3 x^2 -4 x^3 + \ldots). 
\end{equation}

On the other hand, using the same notation, the approximate form of (\ref{Peker}) becomes
\begin{equation}
U_c(x) \approx \delta\left[ C_0 + {C_1\over e^{2 a \beta_0 x}+1} +  
{C_2\over [e^{2a \beta_0 x}+1]^2} \right].
\end{equation}
Expanding the exponentials in Taylor series we get 
\begin{equation}\label{Taylor2}
U_c(x) \approx \delta \left[ \left( C_0 + {C_1\over 2} + {C_2 \over 4}\right) -{\beta_0 \over 4 a_0} (C_1+C_2) x 
+ {\beta_0^2 \over 16 a_0^2} C_2 x^2 +\ldots    \right],
\end{equation}
where
\begin{equation}\label{a0} 
 a_0= {1\over 2a} .
\end{equation}

Equating the coefficients of equal powers of $x$ in  (\ref{Taylor1}) and (\ref{Taylor2}) we get 
\begin{equation}\label{deltaC}
\quad C_0= 1-{4 a_0 \over \beta_0} + {12 a_0^2\over \beta_0^2}, \quad C_1= {8 a_0\over \beta_0} -{48 a_0^2 \over \beta_0^2} , \quad C_2={48 a_0^2\over \beta_0^2}.
\end{equation}

Using the approximate expression of (\ref{Peker}) in  (\ref{WSbarrier}),
we obtain 
\begin{equation}\label{Peker2}
U(\beta) \approx \delta C_0 - {(U_0-\delta C_1)\over e^{2a(\beta-\beta_0)}+1} +  {\delta C_2\over [e^{2a(\beta-\beta_0)}+1]^2},
\end{equation}
where indeed the centrifugal term has been ``absorbed'' in the potential. 

The accuracy of approximating the potential of (\ref{WSbarrier}) by  (\ref{Peker2}) will be discussed in detail 
in section 5, in relation to the parameter values occurring for the nuclei under consideration 
(see figs. 4, 5, 6 and relevant discussion in section 5). 

\subsection{Modified spherically symmetric Woods--Saxon potential}

\begin{figure}[hbt]
\includegraphics[width=0.5\textwidth]{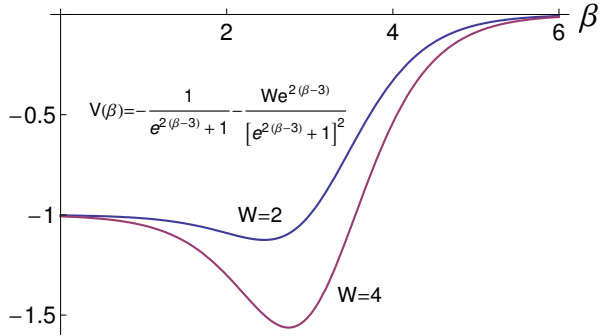}
\caption{(Color online) The Woods--Saxon potential with a dip near its surface, given in (\ref{dip}), is shown
for special values of its free parameters, used for simplicity.
All quantities shown are dimensionless.  }
\end{figure}

On the other hand, the modified spherically symmetric Woods--Saxon potential, which presents a dip
near its surface \cite {Koura}, has the form 
\begin{equation}\label{dip}
V(\beta)= -{V_0\over e^{2a(\beta-\beta_0)}+1}-{W e^{2a(\beta-\beta_0)} \over [e^{2a(\beta-\beta_0)}+1]^2},
\end{equation}
where $V_0$, $W$, $a$ and $\beta_0$ are non-negative free parameters. This potential, shown in figure 2, 
is known \cite{Koksal} to possess
exact solutions of the one-dimensional Schr\"{o}dinger equation 
in terms of Jacobi polynomials \cite{AbrSte}
\begin{equation}
\Psi_n(\beta) \propto \left[ {1\over e^{2a(\beta-\beta_0)}+1} \right]^{b/2} \left[ 1-  {1\over e^{2a(\beta-\beta_0)}+1}\right]^{c/2} P^{(b,c)}_n 
\end{equation}
where 
\begin{equation}\label{bc}
b= {1\over 2} \left( \tilde\rho_n + {V_0\over a^2 \tilde\rho_n} \right), \qquad 
c= {1\over 2} \left( \tilde\rho_n - {V_0\over a^2 \tilde\rho_n} \right),
\end{equation}
with 
\begin{equation}
\tilde\rho_n = \sqrt{1+{W\over a^2}}-(2n+1),
\end{equation}
and the condition \cite{AbrSte} 
 \begin{equation}\label{cond}
 b,c >-1
\end{equation} has to be fulfilled, 
implying the restriction 
\begin{equation}\label{bc2}
b+c > -2. 
\end{equation}

The strengths of the potential of  (\ref{dip}) can be expressed in the form \cite{Koksal} 
\begin{equation}
V_0= a^2 (b^2-c^2) \geq 0, 
\end{equation}
\begin{equation}\label{WW}
W = a^2 [ (b+c) (b+c +4n+2) +4n(n+1)] \geq 0, 
\end{equation}
which imply stronger restrictions on $b$, $c$. 
Indeed, from (\ref{WW}) for $n=0$ one should have $(b+c)(b+c+2) \geq 0$, 
which implies either $b+c \leq -2$, which is not allowed by  (\ref{bc2}), 
or 
\begin{equation}\label{bc0}
b+c \geq  0,
\end{equation}
which is a stronger restriction. 
Higher values of $n$ do not impose any further restrictions, since (\ref{WW}) 
leads to $b+c \leq -2(n+1)$, which is not allowed by (\ref{bc2}), or 
$b+c \geq -2n$, which is weaker than (\ref{bc2}). It should be remembered that the quantity $W$,
given in (\ref{WW}), cannot vanish, since in that case the potential of (\ref{dip}) 
would collapse to the usual Woods--Saxon potential, which is known to have no closed analytical solutions
for its spectrum \cite{Koksal,Flugge}. 

The eigenvalues of the energy are known to be \cite{Koksal} 
\begin{equation}\label{energy}
E_n = -{a^2\over 4} \left(\tilde\rho_n^2 + {V_0^2\over a^4 \tilde\rho_n^2} \right)-{V_0\over 2}.  
\end{equation}

Using the identity  
\begin{equation}
{e^{2a(\beta-\beta_0)} \over [e^{2a(\beta-\beta_0)}+1]^2 } = {1 \over e^{2a(\beta-\beta_0)}+1} 
- {1 \over [e^{2a(\beta-\beta_0)}+1]^2},
\end{equation}
the potential of (\ref{dip}) can be rewritten as
\begin{equation}\label{Veff} 
V(\beta)= -{V_0+W \over e^{2a(\beta-\beta_0)}+1}+{W  \over [e^{2a(\beta-\beta_0)}+1]^2}. 
\end{equation}

\subsection{Spectrum of the modified Woods-Saxon potential with centrifugal barrier}

We remark that this potential has the same form as the potential of (\ref{Peker2}), with  
\begin{equation}
V_0+W = U_0-\delta C_1, \qquad W= \delta C_2,
\end{equation}
leading to
\begin{equation}\label{VW}
V_0 = U_0-{4l(l+1)\over a \beta_0^3}, \qquad W={12 l(l+1)\over a^2 \beta_0^4}.
\end{equation}

Substituting these expressions in  (\ref{energy}) and taking into account the first term in  (\ref{Peker2})
the eigenvalues of the energy for the potential of (\ref{Peker2}) are found to be 
\begin{equation}\label{En}
\fl E_n= {l(l+1)\over \beta_0^2} \left( 1+{12 a_0^2\over \beta_0^2 } \right)
-\left( {\tilde\rho_n \over 4 a_0} \right)^2 - \left({U_0a_0-{8l(l+1)a_0^2\over \beta_0^3} \over \tilde\rho_n} \right)^2 -{U_0\over 2},
\end{equation}
where 
\begin{equation}\label{rho}
\tilde\rho_n= \sqrt{1+{192 l(l+1) a_0^4\over \beta_0^4}}-(2n+1). 
\end{equation}

Therefore we have managed to obtain the energy eigenvalues of the WS potential with a repulsive barrier, 
as approximated through the Pekeris approximation, by exploiting the known solutions of the modified
spherically symmetric WS potential, without having to solve the Schr\"{o}dinger equation anew. 

We remark that the condition of (\ref{bc0}) leads to severe restrictions for $n$.
Indeed, using (\ref{bc0}) we have  
\begin{equation}
b+c = \tilde \rho_n \geq 0,  
\end{equation}
which, using  (\ref{rho}), leads to 
\begin{equation}\label{condit}
n \leq  {1\over 2} \left(- 1+ \sqrt{1 + {192l(l+1)a_0^4 \over \beta_0^4}} \right).
\end{equation} 
Thus in the case of $l=0$, only $n=0$ is allowed. 

This result is valid in a 3-dimensional space. Following the same path 
one can see \cite{Badalov2,Badalov} that the result is also valid in D dimensions,
with the angular momentum $l$ replaced by
\begin{equation}
l_D= l+ {D-3\over 2}. 
\end{equation}
In the special case of a 5-dimensional space, $l$ is replaced by 
\begin{equation}
L=l+1.
\end{equation}

\section{Bohr Hamiltonian for $\gamma$-unstable nuclei}

\subsection{Energy eigenvalues} 

The original Bohr Hamiltonian \cite{Bohr} is

\begin{equation}\label{BohrH}
\eqalign{
\fl H = -{\hbar^2 \over 2B} \left[ {1\over \beta^4} {\partial \over \partial \beta} \beta^4 {\partial \over \partial \beta} + {1\over \beta^2 \sin 3\gamma} {\partial \over \partial \gamma} \sin 3 \gamma {\partial \over \partial \gamma} \right. \\
\left. - {1\over 4 \beta^2} \sum_{k=1,2,3} {Q_k^2 \over \sin^2 \left(\gamma - {2\over 3} \pi k\right) } \right] +V(\beta,\gamma),}
\end{equation}
where $\beta$ and $\gamma$ are the usual collective coordinates describing the 
shape of the nuclear surface,
$Q_k$ ($k=1$, 2, 3) are the components of angular momentum, and $B$ is the 
mass parameter. In what follows we are going to consider $\hbar^2 = 2B =1$. 

Assuming that the potential depends only on the variable $\beta$, 
i.e. $V(\beta,\gamma) = U(\beta)$, one can proceed to separation of variables 
in the standard way \cite{Bohr,Wilets}, using the wavefunction 
\begin{equation} 
\Psi(\beta,\gamma, \theta_i) = {1\over \beta^2} f(\beta) \Phi(\gamma, \theta_i),
\end{equation}  
where $\theta_i$ $(i=1,2,3)$ are the Euler angles describing the orientation 
of the deformed nucleus in space. 

In the equation involving the angles, the eigenvalues of the second order 
Casimir operator of SO(5) occur, having the form 
 $\Lambda = \tau(\tau+3)$, where $\tau=0$, 1, 2, \dots is the quantum 
number characterizing the irreducible representations (irreps) of SO(5), 
called the ``seniority'' \cite{Rakavy}. This equation has been solved 
by B\`es \cite{Bes}.   

The ``radial'' equation yields 
\begin{equation}\label{radial}
-{d^2 f\over d\beta^2} + \left( U(\beta)+ {(\tau+1)(\tau+2)\over \beta^2}  \right) f(\beta)= E_{n,\tau} f(\beta) .
\end{equation}
The effective potential appearing in this equation coincides with that of (\ref{WSbarrier}),
with the formal replacement of $l$ by $\tau+1$. 
Substituting in (\ref{En}) we obtain the energy eigenvalues
\begin{equation}\label{Ene}
\fl E_{n,\tau}= {(\tau+1)(\tau+2)\over \beta_0^2} \left( 1+{12 a_0^2\over \beta_0^2 } \right)
-\left( {\bar\rho_{n,\tau} \over 4 a_0} \right)^2 - \left({U_0 a_0-{8(\tau+1)(\tau+2)a_0^2\over \beta_0^3} \over \bar\rho_{n,\tau}} \right)^2 -{U_0\over 2},
\end{equation}
where 
\begin{equation}
\bar\rho_{n,\tau}= \sqrt{1+{192 (\tau+1)(\tau+2) a_0^4\over \beta_0^4}}-(2n+1). 
\end{equation}

The restriction of (\ref{condit}) in this case reads 
\begin{equation}\label{condit1}
n \leq {1\over 2} \left( -1+ \sqrt{1 + {192(\tau+1)(\tau+2)a_0^4 \over \beta_0^4}} \right).
\end{equation} 
As we shall see in subsection 5.2, only $n=0$ is acceptable for the parameter values occurring in real nuclei.
The violation of the condition of  (\ref{condit1}) even for $n=1$ has as a consequence to push the $n=1$ bands too low in energy,
making them unphysical.  

\subsection{Rescaling the energy eigenvalues}

Eq. (\ref{VW}) takes the form 
\begin{equation}
U_0 = V_0 + {8 a_0 (\tau+1) (\tau+2) \over \beta_0^3},
\end{equation} 
implying that if $V_0$ is a constant, then $U_0$ 
depends on $\tau$, and vice versa. We are going to consider $V_0$ as being the constant quantity. 
Then we can use this equation 
in order to eliminate $U_0$ from (\ref{Ene}). 
Using the notation 
\begin{equation} \label{tildeA}
\tilde A= a_0/\beta_0,
\end{equation} 
the equation for the energy  becomes 
\begin{equation}\label{Espec}
E_{n,\tau} = {(\tau+1)(\tau+2)\over \beta_0^2} (1+12 \tilde A^2-4 \tilde A)
- {\bar\rho_{n,\tau}^2 \over 16 a_0^2} 
-{ V_0^2 a_0^2  \over \bar\rho_{n,\tau}^2}-{V_0\over 2}, 
\end{equation}
with
\begin{equation}\label{barrho}
\bar\rho_{n,\tau} = \sqrt{1+192(\tau+1)(\tau+2) \tilde A^4}-(2n+1).
\end{equation} 
  
In order to reduce the number of free parameters, we fit nuclear spectra leaving out overall scales,
as it is done in the E(5) and X(5) models \cite{IacE5,IacX5}.
Instead of fitting the raw experimental levels $E_L$, where $L$ is the angular momentum, the quantities 
\begin{equation}\label{E20}
{E_L - E_0 \over E_2-E_0}
\end{equation} 
are used. In other words, the energy of the ground state is subtracted from all levels, 
and then each level is divided by the energy of the first excited state,
which in even-even nuclei is the energy of the first excited state with $L=2$.
The connection between $L$ and $\tau$ is described in detail in subsection \ref{spec}.

In the present case this has the following consequences on (\ref{Espec}):

1) The last term, $-V_0/2$, plays no role, since it is a constant and cancels out. 

2) Using the rescaling 
\begin{equation}\label{scal}
 \beta_0 = {\tilde \beta_0 \over \sqrt{V_0}}, \qquad  a_0 = {\tilde a_0 \over \sqrt{V_0}},
\end{equation}
a common factor $V_0$ appears in all terms of (\ref{Espec}), which therefore cancels out when 
(\ref{E20}) above is used.

3) Using $\tilde A= a_0/\beta_0= \tilde a_0 / \tilde \beta_0$, a common factor $\tilde \beta_0^2$ can be taken away from the denominator 
of all terms of  (\ref{Espec}).

Then the equation to be used for the energy fits  becomes 
\begin{equation}\label{Efit1}
\bar E _{n,\tau} = (\tau+1)(\tau+2) (1+12 \tilde A^2-4 \tilde A)- {\bar\rho_{n,\tau}^2\over 16 \tilde A^2} -{\tilde A^2 \tilde \beta_0^4 \over \bar\rho_{n,\tau}^2},
\end{equation}
where $\bar\rho_{n,\tau}$ is still given by (\ref{barrho}). 

This is the equation used in the fits, from which the parameters $\tilde A$ and $\tilde \beta_0$ can be determined. 
Then $\tilde a_0$ is calculated from $\tilde a_0 = \tilde A \tilde \beta_0$. 
 
\subsection{Fixing the scale} 

The Woods--Saxon potential is essentially different from zero within its radius, which in this case is $\beta_0$. 

One way to fix the scale is to identify the radius of the potential, $\beta_0$, 
with the $\beta_{exp}$ value obtained from the experimental values of $B(E2; 0_1^+\to 2_1^+)$
\cite{Raman}. Then from (\ref{scal}) one simply has
\begin{equation}
V_0={\tilde \beta_0^2 \over \beta_{exp}^2}. 
\end{equation} 

Notice that in this case 
$a_0$ can be calculated directly from $a_0= \tilde A \beta_{exp}$.

It should be recalled at this point that while the $B(E2; 0_1^+\to 2_1^+)$ values are experimental 
quantities which are model independent, the deformation parameter is model dependent \cite{Raman}. 
In the present case, a uniform charge distribution out to the distance $R(\theta,\phi)$ is assumed, with zero charge 
beyond it, while the nuclear radius is taken to be $R_0=1.2$ $A^{1/3}$ fm. Then the deformation is connected 
to the transition probability by 
\begin{equation}
\beta = {4\pi \over 3 Z R_0^2} \sqrt{B(E2; 0_1^+\to 2_1^+)\over e^2},
\end{equation}
where $B(E2; 0_1^+\to 2_1^+)$ is measured in $e^2$b$^2$. It should be noticed that $\beta$ is the deformation 
parameter defined in the Bohr framework \cite{Bohr}. A slightly different deformation parameter $\varepsilon$
is used in the framework of the Nilsson model \cite{Nilsson}, connected to the Bohr parameter by 
$\varepsilon= 0.946 \beta$ (see p. 125 of \cite{Nilsson}, or eq. (2.82) of \cite{Ring}).  

\subsection{The spectrum}\label{spec}

The spectrum is characterized by the O(5)$\supset$SO(3) symmetry \cite{Rakavy,Bes}. 
$\tau$ and $L$ are the quantum numbers characterizing the irreps of O(5) and SO(3)
respectively. The values of angular momentum $L$ contained in each 
irrep of O(5) (i.e. for each value of $\tau$) are given by the algorithm \cite{IA} 
\begin{equation}\label{eq:e11} 
\tau = 3 \nu_\Delta +\lambda, \qquad \nu_\Delta =0, 1, \ldots, 
\end{equation}
\begin{equation}\label{eq:e12} 
L=\lambda, \lambda+1, \ldots, 2\lambda -2, 2\lambda 
\end{equation}
(with $2\lambda -1$ missing), 
where $\nu_\Delta$ is the missing quantum number in the reduction 
O(5) $\supset$ SO(3), and are listed in Table 1. 

The ground state band (gsb) has $n=0$ and levels $L_g=0$, 2, 4, 6, \dots, for which $\nu_\Delta=0$ and $L_g = 2\tau$.
Thus within the gsb one has 
\begin{equation}
(\tau+1)(\tau+2)= {(L+2)(L+4)\over 4}. 
\end{equation}

The quasi-$\gamma_1$ band has $n=0$ and levels with $L_\gamma=2$, 3, 4, 5, \dots, 
which are characterized by $\nu_\Delta=0$ and $L_\gamma=2\tau-2$ for $L_\gamma$ being even, 
or by $L_\gamma=2\tau-3$ for $L_\gamma$ being odd. Therefore they exhibit the following degeneracies
\begin{equation}
2_\gamma = 4_g, \quad 3_\gamma=4_\gamma=6_g, \quad 5_\gamma =6_\gamma =8_g, \quad 7_\gamma=8_\gamma =10_g, \ldots   
\end{equation}

The quasi-$\beta_1$ band has levels $L_\beta =0$, 2, 4, 6, \dots. There are two choices:

1) $n=0$, in which the levels have $\nu_\Delta=1$ and $L_\beta=2\tau-6$, leading to
the following degeneracies 
\begin{equation}\label{deg2}
0_\beta = 6_g, \quad 2_\beta=8_g, \quad 4_\beta = 10_g, \quad 6_\beta = 12_g, \ldots 
\end{equation}

2) $n=1$, in which the levels have $\nu_\Delta=0$ and $L_\beta=2\tau$. Then no degeneracies of this kind occur. 

\section{Bohr Hamiltonian for prolate deformed nuclei}

\subsection{Energy eigenvalues} 

If the potential has a minimum around
$\gamma=0$, as it is the case for prolate deformed nuclei,
the angular momentum term in (\ref{BohrH}) can
be written  \cite{IacX5} as
\begin{equation}\label{Sum}
\sum _{k=1}^{3} {Q_k^2 \over \sin^2 \left( \gamma -{2\pi \over 3}
k\right)} \approx {4\over 3} (Q_1^2+Q_2^2+Q_3^2) +Q_3^2 \left(
{1\over \sin^2\gamma} -{4\over 3}\right).
\end{equation}
Exact separation of $\beta$ from the rest of the variables can be obtained for potentials of the form
\cite{Wilets,Fortunato}
\begin{equation}
U(\beta, \gamma) = u(\beta) + {v(\gamma)\over \beta^2}.
\end{equation} 
Concerning the $\gamma$ and Euler angles, the solution provided by B\`es \cite{Bes} 
is not valid any more, thus we seek, as is customary in the literature (see Chapter IV 
of the review article \cite{Fortunato} for further details) wave functions of the form 
\begin{equation}\label{Wavefunction}
\psi(\beta,\gamma,\theta_j)=\xi_L(\beta)\Gamma_K(\gamma){\cal
D}_{M,K}^L(\theta_j),
\end{equation}
where $\theta_j$ ($j=1$, 2, 3) are the Euler
angles, ${\cal D}(\theta_j)$ represents Wigner functions of these
angles, $L$ stands for the eigenvalues of the angular momentum, while $M$ and $K$ are
the eigenvalues of the projections of the angular momentum on the
laboratory-fixed $z$-axis and the body-fixed $z'$-axis respectively.
In these wave functions the dependence on the Euler angles is entering through the Wigner functions, 
as in the many examples reviewed in \cite{Fortunato}, 
while $K$ is assumed to be a good quantum number, an assumption which is not {\it a priori} justified,
since strong $K$ mixing can be present. 

The $\gamma$-equation occurring from this separation of variables has been solved 
in \cite{ESD}, using a potential 
\begin{equation}\label{Vgamma}
v(\gamma)={(3c)^2\gamma^2},
\end{equation}
its eigenfunctions written in terms of Laguerre polynomials 
and its energy eigenvalues given by 
\begin{equation}\label{egam}
\epsilon_\gamma=(3C)(n_\gamma+1), \qquad C=2c, \quad n_\gamma=0,1,2,3,...,
\end{equation}
while the separation constant is 
\begin{equation}\label{lambd}
\lambda= \epsilon_\gamma -{K^2\over 3}.
\end{equation}
It should be noticed at this point that the $\gamma$ potential used in the Bohr Hamiltonian 
has to be periodic in $\gamma$ \cite{Bohr,Baranger,Corrigan}. The potential of (\ref{Vgamma}), used 
in several solutions \cite{IacX5,Fortunato,ESD}  
is a lowest order approximation, representing the first $\gamma$-dependent term in the Taylor expansion 
of the proper periodic potential. There has been recently extensive work involving proper periodic $\gamma$ potentials
in the Bohr Hamiltonian \cite{Fortunato,Buganu,DeBaerde,Gheorghe,Gheorghe2,Buganu39,Buganu40,Buganu88}.  

 The radial equation takes the form \cite{ESD}
\begin{equation} \label{X5Radial}
\left[ -{1\over \beta^4} {\partial \over \partial \beta} \beta^4
{\partial \over \partial \beta} + {L(L+1)\over
3\beta^2}+{\lambda\over\beta^2} +u(\beta) \right] \xi_L(\beta)
 = \epsilon \xi_L(\beta).
\end{equation}
Transforming $\xi_L$ into $\chi_L$ by the relation
\begin{equation}\label{X5Betatransformation}
\xi_L(\beta)= {\chi_L(\beta)\over \beta^2},
\end{equation}
 we obtain
\begin{equation}\label{X5Secondorder}
\chi_L^{''}(\beta)+\left[\epsilon-{{L(L+1)\over3}+\lambda+2\over\beta^2}-u(\beta)\right]\chi_L(\beta)=0.
\end{equation}
We remark that this equation looks very similar to (\ref{radial}),
with ${{L(L+1)\over 3}+\lambda+2}$ replacing $(\tau+1)(\tau+2)$. 
Therefore from (\ref{Ene}) we see that the energy eigenvalues become 
\begin{equation}\label{unscaled2}
\fl E_{n,L}= {\left({{L(L+1)\over 3}+\lambda+2}\right) \over \beta_0^2} \left( 1+{12 a_0^2\over \beta_0^2 } \right)
-\left( {\tilde\rho_{n,L} \over 4 a_0} \right)^2 - \left({U_0 a_0-{8\left({{L(L+1)\over 3}+\lambda+2}\right)a_0^2\over \beta_0^3} \over \tilde\rho_{n,L}} \right)^2 -{U_0\over 2},
\end{equation}
where 
\begin{equation}\label{tilderho}
\tilde\rho_{n,L} = \sqrt{1+192 \left({{L(L+1)\over 3}+\lambda+2}\right)  {a_0^4\over \beta_0^4}}-(2n+1).
\end{equation} 

The restriction of (\ref{condit}) in this case reads 
\begin{equation}\label{condit2}
n \leq  {1\over 2} \left( -1+ \sqrt{1 + 192\left({L(L+1)\over 3}+\lambda+2 \right) {a_0^4 \over \beta_0^4}} \right).
\end{equation} 
As we shall see in subsection 5.1, only $n=0$ is acceptable for the parameter values occurring in real nuclei.
The violation of the condition of  (\ref{condit2}) even for $n=1$ has as a consequence the failure of the relevant
fitting attempts. 

\subsection{Rescaling the energy eigenvalues}

Following the same rescaling procedure as in the $\gamma$-unstable case, the equation used in the fits becomes 
\begin{equation}\label{Efit2}
\bar E _{n,L} = \left({{L(L+1)\over 3}+\lambda+2}\right) (1+12 \tilde A^2-4 \tilde A)
- {\tilde\rho_{n,L}^2\over 16 \tilde A^2} -{\tilde A^2 \tilde \beta_0^4 \over \tilde\rho_{n,L}^2},
\end{equation}
where 
\begin{equation}
\tilde
\rho_{n,L} = \sqrt{1+192 \left({{L(L+1)\over 3}+\lambda+2}\right)  \tilde A^4}-(2n+1).
\end{equation}

This is the equation used in the fits, from which the parameters $\tilde A$, $\tilde \beta_0$, and $C$ (appearing 
in $\lambda$) can be determined. 

\begin{figure*}[hbt]
\includegraphics[width=1\textwidth]{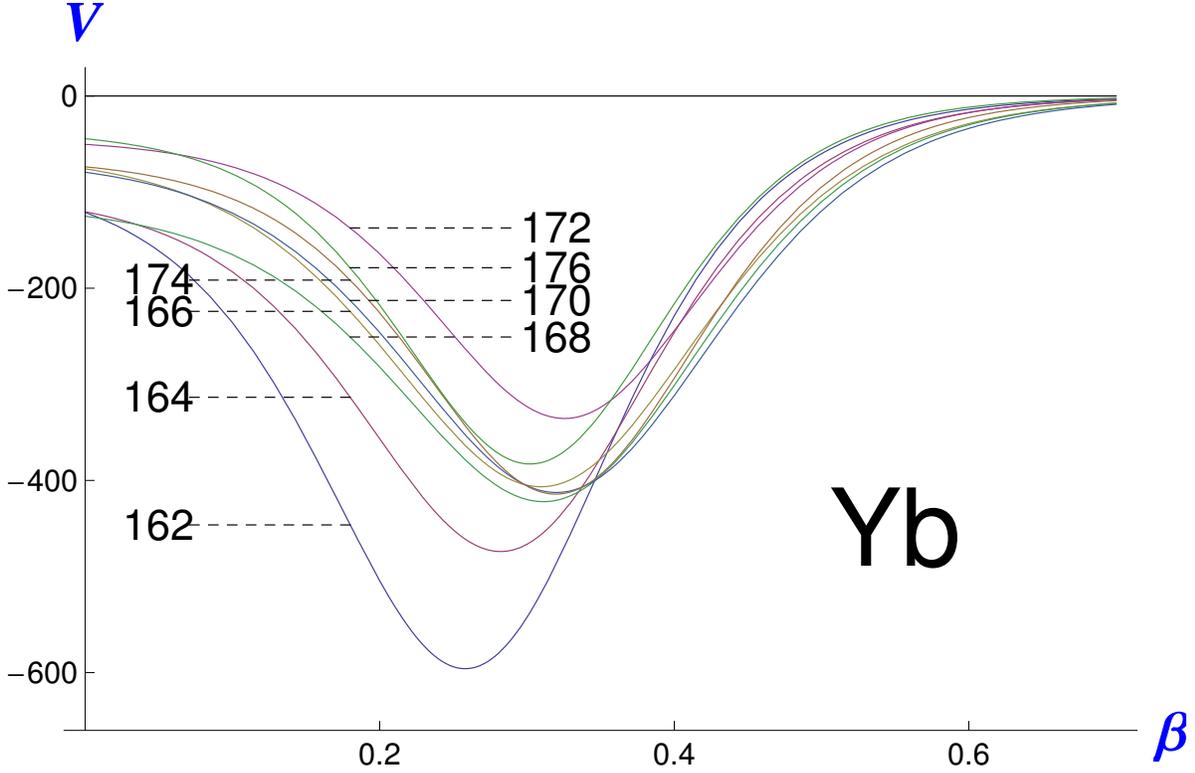}
\caption{(Color online) Effective potentials for $L=10$ for some Yb isotopes, obtained from  (\ref{Veff}).
The parameters are taken from Table \ref{tab1}, while $W$ is given by  (\ref{VW}) with $l(l+1)$ replaced by 
${L(L+1)\over 3}+\lambda+2$, where $\lambda =3C$ from  (\ref{lambd}) and (\ref{egam}). 
The quantities shown are dimensionless. }
\end{figure*}

\section{Numerical results}

\subsection{Prolate deformed nuclei}\label{numresprol}

As a first test, the spectra of 63 nuclei with $R_{4/2} > 2.9 $ have been fitted.
The quality measure
\begin{equation}\label{eq:e99}
\sigma_N = \sqrt{ { \sum_{i=1}^N (E_i(exp)-E_i(th))^2 \over
(N-1)(E(2_1^+))^2} },
\end{equation}
has been used in the rms fits.

The results for 46 nuclei are shown in Table \ref{tab1}. The following comments apply.

1) Only the ground state band and the quasi-$\gamma_1$ band have been included in the fits.

2) When trying to include the quasi-$\beta_1$ band in the fits, trying to correspond it to the $n=1$ case,
the fits fail. The failure is due to violation of the condition of (\ref{condit2}).  

The parameters obtained from fitting the experimental spectra of  $^{162-176}$Yb, shown in Table \ref{tab1}, 
exhibit the following features. 

1) The depth of the potential, $V_0$, exhibits a clear minimum near the middle of the neutron shell ($^{174}_{70}$Yb$_{104}$), 
where maximum deformation is observed, as indicated by the clear maximum exhibited by $\beta_0 \equiv \beta_{exp}$, the experimental values of deformation obtained from the $B(E2; 0^+\to 2^+)$ \cite{Raman}. 

2) The highest diffuseness $\tilde a_0$ occurs nearest to the shell closure, while the lowest appears near the middle of the shell.  

3) In retrospect, the selection of the parameters used in the fits was physically meaningful. $\tilde \beta_0$ is related to the ``well size'' \cite{finite}, while $\tilde A$ is related to the ratio of the diffuseness parameter $a_0$ over the average width of the potential $\beta_0$. 
  
4) The potentials obtained from the Yb isotopes are shown in figure 3, corroborating the above made remarks about maximum depth and diffuseness nearest to the shell closure, and  minimum depth and diffuseness near the middle of the shell. We remark that  the dip near the surface of the modified spherically symmetric WS potential is becoming very large, dominating the overall shape of the potential. It is instructive to compare the shapes of the potentials appearing in figure 3 to these in figure 2, 
since they come from (\ref{Veff}) and (\ref{dip}) respectively, which are equivalent. 

\begin{figure*}[hbt]
\includegraphics[width=1\textwidth]{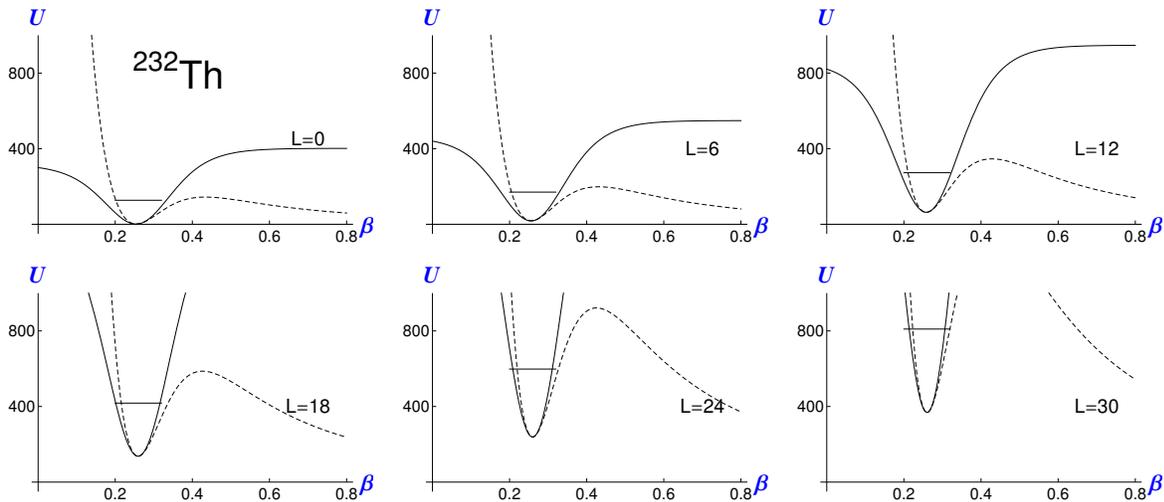}
\caption{(Color online) Exact (dashed lines) and approximate (solid lines) effective potentials for $L=0$, 6, 12, 18, 24, 30 for $^{232}$Th, obtained from  (\ref{WSbarrier}) and  (\ref{Peker2}) respectively.
The parameters are taken from Table \ref{tab1}, providing the quantities needed from  (\ref{delt}), 
(\ref{a0}), and (\ref{deltaC})
while $U_0$ is given by (\ref{VW}) with $l(l+1)$ replaced by ${L(L+1)\over 3}+\lambda+2$, where $\lambda =3C$ from  (\ref{lambd}) and (\ref{egam}). The corresponding energy levels, determined from  (\ref{unscaled2}), are also shown. 
The quantities shown are dimensionless. See subsection \ref{numresprol} for further discussion.}
\end{figure*}

The spectra obtained for some rare earths and actinides are shown in Table \ref{tab2}. Good agreement with the data is obtained.

A question arising at this point is the accuracy of the Pekeris approximation used in the derivation 
of the relevant formulae. In order to check this, we show in figure 4 both the original exact potential
(dashed lines), obtained from (\ref{WSbarrier}), and the approximate potential (solid lines) occurring after the Pekeris approximation,
obtained from (\ref{Peker2}), for six different values of the angular momentum $L$, in the case of $^{232}$Th. In addition,
the relevant energy levels, obtained from (\ref{unscaled2}), are shown. The following comments apply.

1) Near their bottoms, the exact and the approximate potentials nearly coincide. They exhibit minima very close 
to each other, while in addition the two potentials form similar wells around the minima. 

2) At higher energies, the approximate potential wells are slightly displaced to the left, in relation to the exact 
potentials.  However, at the energy of main interest, i.e. at the energy of the relevant level, the widths 
of the two potentials are nearly equal. 

3) The approximation appears to be better at higher values of $L$, i.e. at higher energy levels.
This is expected, since the fits were obtained through an rms procedure. 

\begin{figure*}[hbt]
\includegraphics[width=1\textwidth]{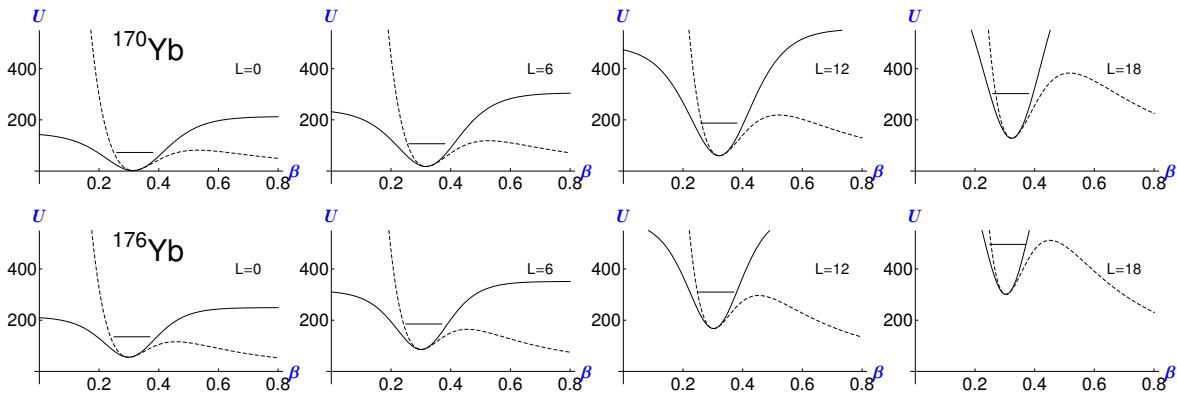}
\caption{(Color online) Same as figure 4, but for $L=0$, 6, 12, 18, for $^{170}$Yb (a) and $^{176}$Yb (b).
See subsection \ref{numresprol} for further discussion.}
\end{figure*}

The same observations hold for all nuclei shown in Table \ref{tab1}, with two exceptions, $^{172}$Yb and $^{176}$Yb, 
the latter exhibited in figure 5. 
While $^{170}$Yb (and the rest of the Yb isotopes shown in Table \ref{tab1}) exhibit the qualitative behaviour 
described above, in $^{176}$Yb (as  well as in $^{172}$Yb) we remark that the approximation breaks down, since for the lowest values of $L$
the relevant energy level lies higher than the corresponding barrier on the rhs of the exact potential. 
In other words, we still get fits of good quality, which are obtained using the approximate potential,
but this potential is not similar to the exact potential any more, preventing us to draw any conclusions 
regarding the exact potential. 
It should be noticed at this point that the approximate potentials (solid lines) for Yb isotopes shown in figures 3 and 5 are identical, up to a displacement by a constant. Indeed, the potentials of figure 3 come from (\ref{Veff}), while the 
potentials of figure 5 come from  (\ref{Peker2}), the latter being displaced in relation to the former by the term
$\delta C_0$. This term is included in figure 5, since it is necessary for the comparison to the exact potentials (dashed lines)
given by (\ref{WSbarrier}).  

$^{172}$Yb and $^{176}$Yb are not isolated cases. Similar breakdowns of the approximation have been found in 17 more nuclei, shown in Table \ref{tab3}. In all cases, good fits of quality comparable to those shown in Table \ref{tab1} are obtained, 
but they correspond to approximate potentials being very different from the exact ones. 

From the numerical results obtained for all the nuclei mentioned above, we see that the breakdown of the approximation 
occurs for ``well size'' $\tilde \beta_0 \leq 1.8$ and/or diffuseness $\tilde a_0 \leq 0.43$. Since $\beta_0^2$ appears in the denominator of the rotational term, this result is understandable. The lower $\beta_0$ becomes, the more important the rotational term, in which the Pekeris approximation has been used, becomes, thus the approximation to the exact potential deteriorates.

We therefore conclude that good fits are obtained for all nuclei studied, but only if the 
``well size'' $\tilde \beta_0$ fulfils the condition $\tilde \beta_0 \geq 1.9$ and the diffuseness 
fulfils the condition $\tilde a_0 \geq 0.44$ the approximate potentials are quite similar to the corresponding exact potentials. 

\begin{figure*}[hbt]
\includegraphics[width=1\textwidth]{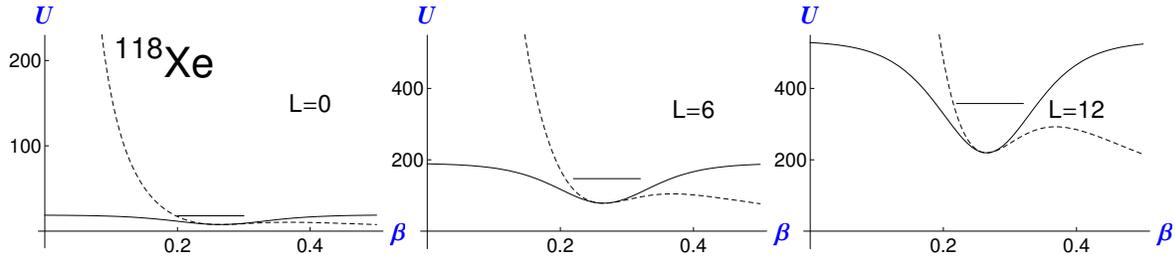}
\caption{(Color online)  Exact (dashed lines) and approximate (solid lines) effective potentials for $\tau=0$, 3, 6
($L=0$, 6, 12)  for 
$^{118}$Xe, obtained from  (\ref{WSbarrier}) and (\ref{Peker2}) respectively.
The parameters are taken from Table \ref{tab4}, providing the quantities needed from  (\ref{delt}), 
(\ref{a0}), and (\ref{deltaC}),
while $W$ is given by (\ref{VW}) with $l(l+1)$ replaced by $(\tau+1)(\tau+2)$. The corresponding energy levels, 
determined from (\ref{Espec}), are also shown. 
The quantities shown are dimensionless. See subsection \ref{numresgunst} for further discussion.  }
\end{figure*}

\subsection{$\gamma$-unstable nuclei}\label{numresgunst}

The spectra of 38 nuclei with $R_{4/2} < 2.6 $ [$^{98,100,102}$Ru, $^{102,104,106,108,110,112,114,116}$Pd, 
$^{108,110,112}$Cd, $^{118,120,122,124,126,128,130,132,134}$Xe, $^{130,132,134}$Ba, $^{134,138}$Ce, $^{142}$Gd, 
$^{156}$Er, $^{186,188,190,192,194,196,198,200}$Pt]
have been fitted, using the same quality measure.

The following comments apply.

1) Only the ground state band and the quasi-$\gamma_1$ band have been included in the fits.

2) When trying to include the quasi-$\beta_1$ band in the fits, trying to correspond it to the $n=1$ case,
one finds it lying too low in energy, due to the violation of the restriction of  (\ref{condit1}). 

3) When trying to include the quasi-$\beta_1$ band in the fits, trying to correspond it to the $n=0$ case,
the levels of the $\beta_1$-band in most cases are overestimated. There is nothing to be done against this, 
since the degeneracies of (\ref{deg2}) fix the position of the theoretical levels. 
Thus this failure is due to the fact that the experimental data do not exhibit exactly the O(5) degeneracies.

The parameters obtained from fitting the experimental spectra of  $^{118-134}$Xe, shown in Table \ref{tab4}.
We remark that in all cases the ``well size'' $\tilde \beta_0$ is much lower than 1.8, while in addition the 
diffuseness is much lower than 0.43~.  The same holds for all of the 38 nuclei considered. 
Therefore the approximate potentials bear little relevance to the exact potentials. 

As an example, we show in figure 6 both the original exact potential
(broken lines), obtained from (\ref{WSbarrier}), and the approximate potential (solid lines) occurring after the Pekeris approximation, obtained from (\ref{Peker2}), for three different values of the seniority quantum number $\tau$, 
in the case of $^{118}$Xe. In addition, the relevant energy levels, obtained from eq. (\ref{Espec}), are shown. 
The following comments apply.

1) Near their bottoms, the exact and the approximate potentials nearly coincide.

2) The approximation breaks down, since for all values of $\tau$
the relevant energy level lies higher than the relevant barrier on the rhs of the exact potential. 
We still get fits of good quality, which are obtained using the approximate potential,
but this potential is not similar to the exact potential any more, preventing us to draw any conclusions 
regarding the exact potential. 

We therefore conclude that although we could derive analytical expressions for fitting the spectra,
the parameter values coming out from the fits correspond to ``well sizes'' and diffuseness for which the approximate 
potential is not quite similar to the exact potential, i.e. the Pekeris approximation breaks down. 

\section{Conclusions}

The main results are summarized here.

1) Approximate solutions in closed form are obtained for the 5-dimensional Bohr Hamiltonian 
with the Woods--Saxon potential, using the Pekeris approximation and the exact solutions of an extended Woods--Saxon potential in one dimension, featuring a dip near its surface.

2) Applying the results to several $\gamma$-unstable and prolate deformed nuclei, 
we find that the WS potential can describe the ground state bands and the $\gamma_1$ bands equally well 
as other potentials (Davidson, Kratzer, Morse), if the ``well size'' and the diffuseness are large enough (at least 1.9
and 0.44 respectively), 
but it fails to describe the $\beta_1$ bands, apparently because of its lack of a hard core. 
Several (forty-four) examples of deformed nuclei satisfying this condition have been found, but on the other hand
all $\gamma$-unstable nuclei considered violate this condition. 

3) The form of the potentials coming out
from the fits exhibits a very large dip near the surface. In other words, the Bohr equation forces 
the parameters of the WS potential to obtain values producing a very large dip near its surface,
so that its overall shape around its minimum largely resembles the shape around the minimum of the Davidson, 
or the Kratzer, or the Morse potential. 

4) The present results suggest that potentials used in the Bohr Hamiltonian 
can provide satisfactory results for nuclear spectra if they possess two features, 
a hard core and a deep oscillator-like minimum. They also suggest that the lack of a hard core does not decisively affect the description
of the ground state and $\gamma_1$ bands, but destroys the ability of the potential to describe $\beta_1$ bands.  

Concerning the position of the quasi-$\beta_1$ bands, which are not reproduced by the Woods--Saxon potential, 
the following comments apply. 

1) The quasi-$\beta_1$ bandheads move to higher energies (normalized to the energy of the first excited state)
both in $\gamma$-unstable and in deformed nuclei, if the left wall of the infinite well potential used 
in the E(5) and X(5) critical point symmetries is gradually moved to the right of the origin of the $\beta$-axis,
approaching the right wall \cite{CBS,CBS2}. 

2) The interlevel spacings within the $\beta_1$- band, which are known to be overestimated 
in the framework of the X(5) critical point symmetry, get fixed by allowing the right wall 
of the infinite well potential to be sloped to the right \cite{sloped}. It should be mentioned here 
that the identification of a symmetry underlying the X(5) special solution of the Bohr Hamiltonian 
remains an open problem.  

Analytical wave functions for the Bohr Hamiltonian with the WS potential can be readily obtained
by exploiting the similarity of this Hamiltonian, after using the Pekeris approximation, to the
exactly soluble extended WS potential with a dip near its surface. The calculation of B(E2) transition rates 
becomes then a straightforward task, to be addressed in further work.  

Finally, the solution of the Bohr Hamiltonian with a Woods--Saxon potential within the framework of the algebraic 
collective model \cite{Rowe735,Rowe753,Caprio,Welsh} would offer the opportunity of comparison of the present approximate results to exact numerical solutions. 

\ack{B. G\"{o}n\"{u}l and M. \c{C}apak acknowledge financial support by the Scientific and Technical Research Council of Turkey (T\"{U}B\.{I}TAK) under project number ARDEB/1002-113F218. D. Petrellis acknowledges financial support by
the Scientific Research Projects Coordination Unit of Istanbul University under Project No 50822.}

 \newpage 

\begin{table}

\caption{Quantum numbers appearing in the O(5)$\supset$SO(3) reduction 
\cite{IA}, occurring from  (\ref{eq:e11}) and (\ref{eq:e12}).  
}

\bigskip

\begin{tabular}{l l l l}
\hline
$\tau$ & $\nu_\Delta$ & $\lambda$ & $L$         \\
       &              &           &             \\  
0      &  0           &     0     &  0          \\
       &              &           &             \\
1      &  0           &     1     &  2          \\
       &              &           &             \\
2      &  0           &     2     &  4,2        \\
       &              &           &             \\
3      &  0           &     3     &  6,4,3      \\
3      &  1           &     0     &  0          \\
       &              &           &             \\
4      &  0           &     4     &  8,6,5,4    \\
4      &  1           &     1     &  2          \\
       &              &           &             \\  
5      &  0           &     5     & 10,8,7,6,5  \\
5      &  1           &     2     & 4,2         \\
       &              &           &             \\
6      &  0           &     6     & 12,10,9,8,7,6\\
6      &  1           &     3     & 6,4,3       \\
6      &  2           &     0     & 0           \\
\hline 
\end{tabular}
\end{table}

\newpage

\begin{table}
\caption{\label{tab1} Comparison of theoretical predictions of the Bohr Hamiltonian 
with the Woods--Saxon potential for 46 axially symmetric prolate deformed nuclei 
to experimental data \cite{ENSDF} of rare earth and actinides with $R_{4/2}$ $>$ 2.9 
and known $0_2^+$ and $2_{\gamma}^+$ states. 
The angular momenta of the highest levels of the ground state
and $\gamma_1$ bands included in the rms fit are labelled by $L_g$
and $L_\gamma$ respectively, while $N$ indicates the
total number of levels involved in the fit and $\sigma$ is the
quality measure of (\ref{eq:e99}). All energies are normalized 
to the energy of the first excited state, $E(2^+_1)$. 
For each band, the $R_{4/2}=E(4_1^+)/E(2_1^+)$ ratio (labelled by $4/2$),
the normalized bandhead of the $\gamma_1$ band (labelled as $2_\gamma/2$), 
and the normalized last members of the ground state and $\gamma_1$
 bands included in the fit 
(labelled by $L_g/2$ and $L_\gamma/2$ respectively), are reported. 
$\beta_0$ has been obtained from \cite{Raman}.
The theoretical predictions are obtained from (\ref{Efit2}).
See subsection \ref{numresprol} for further discussion.}

\tiny

\begin{tabular}{@{} l l l l l l l l l l l l l l l l l l l l}
\br
nucleus & $\tilde \beta_0$ & $\tilde A$ & $C$ & $\tilde a_0$ & $\beta_0$ & $V_0$ & $10^3 a_0$ & $L_g$ & $L_\gamma$ & $N$ & $\sigma$ & 
$4/2$ & $4/2$ & $L_g/2$ & $L_g/2$ & $2_\gamma/2$ & $2_\gamma/2$ & $L_\gamma/2$ & $L_\gamma/2$ \\
 & & & & & & & & & & & & exp & th & exp & th & exp & th & exp & th \\
\mr
$^{150}$Nd & 4.1 & 0.15 & 7.6 & 0.62 & 0.285 &207 & 43 & 14 & 4 &  9 & 0.24 & 2.93 & 3.02 & 20.6 & 20.7 &  8.2 &  8.6 & 10.4 &  9.9 \\
 
$^{152}$Sm & 4.8 & 0.20 & 8.6 & 0.96 & 0.306 &245 & 61 & 16 & 9 & 15 & 0.60 & 3.01 & 3.10 & 27.6 & 28.0 &  8.9 & 10.2 & 19.5 & 18.4 \\
$^{154}$Sm & 4.2 & 0.22 &14.0 & 0.92 & 0.341 &152 & 75 & 16 & 7 & 13 & 0.53 & 3.25 & 3.26 & 36.2 & 36.5 & 17.6 & 18.6 & 26.3 & 25.0 \\

$^{154}$Gd & 3.4 & 0.24 & 6.9 & 0.82 & 0.312 &119 & 75 & 26 & 7 & 18 & 0.35 & 3.02 & 3.14 & 57.3 & 57.4 &  8.1 &  8.8 & 14.7 & 14.2 \\
$^{156}$Gd & 2.2 & 0.24 &10.8 & 0.53 & 0.338 & 42 & 81 & 26 &16 & 27 & 0.68 & 3.24 & 3.25 & 74.0 & 74.1 & 13.0 & 14.5 & 44.9 & 44.0 \\
$^{158}$Gd & 2.8 & 0.18 &10.6 & 0.50 & 0.348 & 65 & 63 & 12 & 6 & 10 & 0.08 & 3.29 & 3.28 & 23.5 & 23.5 & 14.9 & 15.1 & 20.4 & 20.3 \\

$^{156}$Dy & 3.5 & 0.24 & 6.0 & 0.84 & 0.293 &143 & 70 & 28 &13 & 25 & 0.49 & 2.93 & 3.08 & 57.9 & 57.9 &  6.5 &  7.5 & 23.8 & 22.9 \\
$^{158}$Dy & 2.4 & 0.24 & 7.7 & 0.58 & 0.326 & 54 & 78 & 28 & 8 & 20 & 0.46 & 3.21 & 3.21 & 75.4 & 74.8 &  9.6 & 10.3 & 19.1 & 18.2 \\
$^{160}$Dy & 3.0 & 0.23 & 9.3 & 0.69 & 0.339 & 78 & 78 & 24 &23 & 33 & 0.88 & 3.27 & 3.23 & 65.1 & 66.5 & 11.1 & 12.5 & 68.2 & 69.4 \\

$^{160}$Er & 3.0 & 0.24 & 5.6 & 0.72 & 0.304 & 97 & 73 & 26 & 5 & 16 & 0.52 & 3.10 & 3.11 & 55.9 & 55.4 &  6.8 &  7.3 & 10.5 & 10.0 \\
$^{162}$Er & 1.9 & 0.24 & 7.4 & 0.46 & 0.322 & 35 & 77 & 20 &12 & 20 & 0.89 & 3.23 & 3.22 & 43.7 & 45.4 &  8.8 & 10.1 & 28.5 & 27.0 \\
$^{164}$Er & 2.1 & 0.21 & 6.9 & 0.44 & 0.333 & 40 & 70 & 22 &16 & 25 & 0.61 & 3.28 & 3.25 & 61.8 & 62.8 &  9.4 &  9.8 & 41.6 & 43.1 \\

$^{162}$Yb & 2.4 & 0.25 & 4.0 & 0.60 & 0.263 & 83 & 66 & 24 & 4 & 14 & 0.33 & 2.92 & 3.05 & 39.9 & 39.1 &  4.8 &  5.3 &  7.1 &  6.9 \\ 
$^{164}$Yb & 2.9 & 0.23 & 5.6 & 0.67 & 0.290 &100 & 67 & 18 & 5 & 12 & 0.30 & 3.13 & 3.14 & 35.6 & 35.6 &  7.0 &  7.5 & 10.9 & 10.3 \\
$^{166}$Yb & 2.4 & 0.23 & 7.0 & 0.55 & 0.315 & 58 & 72 & 24 &13 & 23 & 0.82 & 3.23 & 3.22 & 62.3 & 63.9 &  9.1 &  9.6 & 31.2 & 30.0 \\
$^{168}$Yb & 3.4 & 0.22 & 8.7 & 0.75 & 0.322 &111 & 71 & 34 & 7 & 22 & 0.64 & 3.27 & 3.22 &120.5 &119.7 & 11.2 & 11.6 & 18.5 & 17.9 \\ 
$^{170}$Yb & 2.6 & 0.22 &10.1 & 0.57 & 0.326 & 64 & 72 & 20 &14 & 22 & 0.63 & 3.29 & 3.26 & 52.7 & 54.1 & 13.6 & 14.0 & 39.3 & 39.9 \\
$^{172}$Yb & 2.2 & 0.19 &13.1 & 0.42 & 0.330 & 44 & 63 & 14 & 5 & 10 & 0.11 & 3.31 & 3.30 & 32.0 & 32.1 & 18.6 & 18.8 & 22.6 & 22.4 \\
$^{174}$Yb & 2.6 & 0.20 &15.1 & 0.52 & 0.325 & 64 & 65 & 18 & 5 & 12 & 0.13 & 3.31 & 3.30 & 50.2 & 50.4 & 21.4 & 21.5 & 25.2 & 25.0 \\
$^{176}$Yb & 1.8 & 0.20 &10.7 & 0.36 & 0.305 & 35 & 61 & 18 & 5 & 12 & 0.29 & 3.31 & 3.29 & 48.4 & 49.0 & 15.4 & 15.3 & 19.0 & 18.9 \\

$^{166}$Hf & 2.2 & 0.25 & 4.1 & 0.55 & 0.250 & 77 & 63 & 22 & 3 & 12 & 0.24 & 2.97 & 3.08 & 36.9 & 36.5 &  5.1 &  5.5 &  6.3 &  6.2 \\
$^{168}$Hf & 2.5 & 0.24 & 5.7 & 0.60 & 0.275 & 83 & 66 & 22 & 4 & 13 & 0.28 & 3.11 & 3.16 & 46.5 & 46.9 &  7.1 &  7.7 &  9.8 &  9.4 \\ 
$^{170}$Hf & 3.9 & 0.22 & 7.9 & 0.86 & 0.301 &168 & 66 & 34 & 4 & 19 & 0.48 & 3.19 & 3.16 &105.7 &105.3 &  9.5 & 10.1 & 12.2 & 11.7 \\
$^{172}$Hf & 3.4 & 0.23 & 9.0 & 0.78 & 0.276 &152 & 63 & 38 & 6 & 23 & 0.69 & 3.25 & 3.21 &132.8 &133.5 & 11.3 & 11.9 & 17.0 & 16.3 \\
$^{174}$Hf & 3.1 & 0.23 &10.2 & 0.71 & 0.286 &117 & 66 & 22 & 4 & 13 & 0.43 & 3.27 & 3.24 & 58.2 & 58.9 & 13.5 & 13.7 & 15.9 & 15.5 \\
$^{176}$Hf & 2.9 & 0.22 &11.4 & 0.64 & 0.295 & 96 & 65 & 18 & 6 & 13 & 0.34 & 3.28 & 3.27 & 45.4 & 45.7 & 15.2 & 15.7 & 21.1 & 20.6 \\
$^{178}$Hf & 2.2 & 0.22 & 9.2 & 0.48 & 0.280 & 62 & 62 & 18 & 6 & 13 & 0.41 & 3.29 & 3.26 & 44.2 & 45.0 & 12.6 & 12.9 & 18.1 & 17.8 \\

$^{176}$W  & 2.4 & 0.24 & 7.6 & 0.58 &       &    &    & 22 & 5 & 14 & 0.31 & 3.22 & 3.21 & 51.8 & 51.8 &  9.6 & 10.2 & 14.0 & 13.2 \\
$^{178}$W  & 2.4 & 0.23 & 7.6 & 0.55 &       &    &    & 14 & 2 &  7 & 0.14 & 3.24 & 3.23 & 27.0 & 27.2 & 10.5 & 10.4 & 10.5 & 10.4 \\
$^{180}$W  & 2.7 & 0.24 & 8.7 & 0.65 & 0.254 &113 & 61 & 24 & 7 & 17 & 0.67 & 3.26 & 3.22 & 60.0 & 60.4 & 10.8 & 11.6 & 18.7 & 17.6 \\

$^{176}$Os & 4.0 & 0.23 & 6.0 & 0.92 &       &    &    & 24 & 5 & 15 & 0.37 & 2.93 & 3.04 & 45.5 & 45.5 &  6.4 &  7.3 & 10.4 &  9.6 \\ 
$^{178}$Os & 3.8 & 0.23 & 6.0 & 0.87 &       &    &    & 16 & 5 & 11 & 0.40 & 3.02 & 3.06 & 26.1 & 26.0 &  6.6 &  7.4 & 10.8 &  9.9 \\
$^{180}$Os & 2.5 & 0.25 & 5.9 & 0.63 & 0.226 &122 & 57 & 14 & 7 & 12 & 0.59 & 3.09 & 3.14 & 21.8 & 22.0 &  6.6 &  7.7 & 14.2 & 13.0 \\
$^{182}$Os & 3.7 & 0.24 & 6.7 & 0.89 & 0.234 &250 & 56 & 26 & 7 & 18 & 0.84 & 3.15 & 3.10 & 54.0 & 53.3 &  7.0 &  8.3 & 14.6 & 13.4 \\ 
$^{184}$Os & 2.2 & 0.24 & 6.2 & 0.53 & 0.213 &107 & 51 & 22 & 6 & 15 & 1.05 & 3.20 & 3.19 & 47.9 & 49.4 &  7.9 &  8.4 & 13.5 & 12.8 \\

$^{228}$Ra & 3.1 & 0.24 &10.1 & 0.74 & 0.217 &204 & 52 & 22 & 3 & 12 & 0.18 & 3.21 & 3.23 & 53.6 & 54.0 & 13.3 & 13.3 & 14.1 & 14.0 \\

$^{228}$Th & 3.5 & 0.24 &13.1 & 0.84 & 0.230 &231 & 55 & 18 & 5 & 12 & 0.15 & 3.24 & 3.25 & 41.7 & 41.6 & 16.8 & 17.1 & 20.3 & 20.0 \\ 
$^{230}$Th & 2.8 & 0.22 &10.6 & 0.62 & 0.244 &132 & 54 & 18 & 4 & 11 & 0.12 & 3.27 & 3.26 & 45.1 & 45.3 & 14.7 & 14.7 & 16.6 & 16.6 \\
$^{232}$Th & 2.4 & 0.23 &12.1 & 0.55 & 0.261 & 85 & 60 & 30 &12 & 25 & 0.54 & 3.28 & 3.27 &104.6 &105.9 & 15.9 & 16.6 & 36.5 & 35.6 \\
 
$^{232}$U  & 3.0 & 0.22 &13.3 & 0.66 & 0.264 &129 & 58 & 20 & 4 & 12 & 0.20 & 3.29 & 3.28 & 55.9 & 56.3 & 18.2 & 18.4 & 20.4 & 20.3 \\
$^{234}$U  & 2.7 & 0.23 &15.9 & 0.62 & 0.272 & 99 & 63 & 28 & 7 & 19 & 0.36 & 3.30 & 3.29 & 98.8 & 99.6 & 21.3 & 21.8 & 29.0 & 28.5 \\
$^{236}$U  & 3.3 & 0.22 &15.5 & 0.73 & 0.282 &137 & 62 & 26 & 5 & 16 & 0.63 & 3.30 & 3.29 & 89.3 & 90.6 & 21.2 & 21.4 & 24.9 & 24.7 \\
$^{238}$U  & 2.9 & 0.23 &18.3 & 0.67 & 0.286 &103 & 66 & 30 &27 & 40 & 0.85 & 3.30 & 3.29 &112.1 &114.9 & 23.6 & 25.0 &112.7 &113.5 \\

$^{240}$Pu & 2.2 & 0.22 &19.0 & 0.48 & 0.289 & 58 & 64 & 26 & 4 & 15 & 0.21 & 3.31 & 3.30 & 95.5 & 96.0 & 26.6 & 26.6 & 28.8 & 28.6 \\
$^{242}$Pu & 2.3 & 0.22 &17.7 & 0.51 & 0.292 & 62 & 64 & 26 & 2 & 13 & 0.51 & 3.31 & 3.30 & 93.7 & 94.8 & 24.7 & 24.8 & 24.7 & 24.8 \\

$^{248}$Cm & 2.7 & 0.22 &17.2 & 0.59 & 0.297 & 83 & 65 & 28 & 2 & 14 & 0.87 & 3.31 & 3.30 & 10.5 & 10.7 & 24.2 & 24.0 & 24.2 & 24.0 \\

\br
\end{tabular}

\end{table}

\newpage

\begin{table}

\caption{\label{tab2}Spectra of some rare earths and actinides, obtained with the parameters shown in Table \ref{tab1}.}

\begin{tabular}{@{} l l l l l l l l l}
\br
     & $^{156}$Gd & $^{156}$Gd & $^{164}$Er & $^{164}$Er & $^{170}$Yb & $^{170}$Yb & $^{232}$Th & $^{232}$Th\\
 $L$ & exp        & th         & exp        & th         & exp        & th         & exp        & th        \\

\mr
               gsb &       &       &       &       &       &       &       &       \\
                 4 &  3.24 &  3.25 &  3.28 &  3.25 &  3.29 &  3.26 &  3.28 &  3.27 \\
                 6 &  6.57 &  6.59 &  6.72 &  6.60 &  6.80 &  6.66 &  6.75 &  6.69 \\
                 8 & 10.85 & 10.86 & 11.21 & 10.92 & 11.43 & 11.07 & 11.28 & 11.13 \\
                10 & 15.92 & 15.88 & 16.61 & 16.11 & 17.06 & 16.37 & 16.75 & 16.47 \\
                12 & 21.63 & 21.56 & 22.79 & 22.09 & 23.54 & 22.49 & 23.03 & 22.60 \\
                14 & 27.83 & 27.78 & 29.58 & 28.83 & 30.63 & 29.36 & 30.04 & 29.45 \\
                16 & 34.39 & 34.50 & 37.33 & 36.29 & 37.92 & 36.94 & 37.65 & 36.97 \\
                18 & 41.29 & 41.66 & 45.10 & 44.44 & 45.18 & 45.21 & 45.84 & 45.11 \\
                20 & 48.62 & 49.23 & 53.28 & 53.27 & 52.66 & 54.13 & 54.52 & 53.85 \\
                22 & 56.49 & 57.18 & 61.85 & 62.77 &       &       & 63.69 & 63.16 \\
                24 & 64.95 & 65.49 &       &       &       &       & 73.32 & 73.03 \\
                26 & 73.99 & 74.15 &       &       &       &       & 83.38 & 83.45 \\
                28 &       &       &       &       &       &       & 93.82 & 94.39 \\
                30 &       &       &       &       &       &       &104.56 &105.86 \\
                   &       &       &       &       &       &       &       &       \\
        $\gamma_1$ &       &       &       &       &       &       &       &       \\
                 2 & 12.97 & 14.49 &  9.41 &  9.80 & 13.60 & 14.02 & 15.91 & 16.59 \\
                 3 & 14.03 & 15.27 & 10.36 & 10.65 & 14.54 & 14.85 & 16.80 & 17.41 \\
                 4 & 15.24 & 16.29 & 11.58 & 11.77 & 15.78 & 15.95 & 18.03 & 18.49 \\
                 5 & 16.94 & 17.55 & 13.10 & 13.15 & 17.33 & 17.31 & 19.45 & 19.82 \\
                 6 & 18.47 & 19.04 & 14.87 & 14.78 & 19.01 & 18.92 & 21.27 & 21.41 \\
                 7 & 20.79 & 20.74 & 16.91 & 16.65 & 21.13 & 20.78 & 23.21 & 23.23 \\
                 8 & 22.61 & 22.65 & 19.09 & 18.76 & 23.19 & 22.87 & 25.50 & 25.29 \\
                 9 & 25.29 & 24.74 & 21.64 & 21.08 & 25.76 & 25.18 & 27.75 & 27.57 \\
                10 & 27.45 & 27.02 & 23.90 & 23.63 & 28.16 & 27.72 & 30.62 & 30.06 \\
                11 & 30.20 & 29.47 & 27.13 & 26.38 & 30.90 & 30.47 & 33.22 & 32.75 \\
                12 & 32.85 & 32.08 & 29.91 & 29.33 & 33.55 & 33.42 & 36.48 & 35.64 \\
                13 & 35.69 & 34.84 & 33.13 & 32.48 & 36.40 & 36.57 &       &       \\
                14 & 38.64 & 37.74 & 35.75 & 35.83 & 39.25 & 39.91 &       &       \\
                15 & 41.76 & 40.78 & 38.51 & 39.36 &       &       &       &       \\
                16 & 44.90 & 43.95 & 41.59 & 43.07 &       &       &       &       \\
                   &       &       &       &       &       &       &       &       \\
\br
\end{tabular}
\end{table}

\newpage

\begin{table}

\caption{\label{tab3}Same as Table \ref{tab1}, for 17 additional nuclei.
See subsection \ref{numresprol} for further discussion. }

\tiny

\begin{tabular}{@{} l l l l l l l l l l l l l l l l l l l l}
\br
 & & & & & & & & & & & & & & & & & & &  \\
nucleus & $\tilde \beta_0$ & $\tilde A$ & $C$ & $\tilde a_0$ & $\beta_0$ & $V_0$ & $10^3 a_0$ & $L_g$ & $L_\gamma$ & $N$ & $\sigma$ &  
$4/2$ & $4/2$ & $L_g/2$ & $L_g/2$ & $2_\gamma/2$ & $2_\gamma/2$ & $L_\gamma/2$ & $L_\gamma/2$ \\
 & & & & & & & & & & & & exp & th & exp & th & exp & th & exp & th \\
\mr
$^{160}$Gd & 1.8 & 0.18 & 9.0 & 0.32 & 0.353 & 26 & 64 & 16 & 8 & 14 & 0.10 & 3.30 & 3.29 & 40.0 & 40.2 & 13.1 & 13.1 & 22.8 & 22.8 \\
$^{162}$Gd & 1.5 & 0.18 & 8.2 & 0.27 &       &    &    & 14 & 4 &  9 & 0.05 & 3.29 & 3.29 & 31.4 & 31.5 & 12.0 & 12.0 & 14.1 & 14.1 \\

$^{162}$Dy & 1.9 & 0.19 & 7.7 & 0.36 & 0.343 & 31 & 65 & 18 &14 & 21 & 0.23 & 3.29 & 3.28 & 47.6 & 47.9 & 11.0 & 11.1 & 39.4 & 39.2 \\
$^{164}$Dy & 1.2 & 0.19 & 7.1 & 0.23 & 0.348 & 12 & 66 & 20 &10 & 18 & 0.26 & 3.30 & 3.28 & 57.4 & 58.1 & 10.4 & 10.4 & 25.3 & 25.3 \\
$^{166}$Dy & 1.2 & 0.15 & 7.5 & 0.18 &       &    &    &  6 & 5 &  6 & 0.02 & 3.31 & 3.30 &  6.9 &  6.8 & 11.2 & 11.2 & 14.9 & 14.9 \\

$^{166}$Er & 1.2 & 0.21 & 6.9 & 0.25 & 0.342 & 12 & 72 & 16 &14 & 20 & 0.30 & 3.29 & 3.26 & 36.8 & 37.1 &  9.8 &  9.9 & 35.7 & 36.6 \\ 
$^{168}$Er & 1.2 & 0.16 & 6.9 & 0.19 & 0.338 & 13 & 54 & 18 & 8 & 15 & 0.20 & 3.31 & 3.30 & 50.0 & 50.6 & 10.3 & 10.3 & 20.4 & 20.3 \\
$^{170}$Er & 0.7 & 0.18 & 8.9 & 0.13 & 0.336 &  4 & 61 & 26 &19 & 30 & 0.85 & 3.31 & 3.30 & 95.8 & 98.1 & 11.9 & 13.0 & 66.2 & 65.7 \\

$^{178}$Yb & 1.3 & 0.17 & 9.9 & 0.22 &       &    &    &  6 & 2 &  3 & 0.01 & 3.31 & 3.30 &  6.9 &  6.9 & 14.5 & 14.5 & 14.5 & 14.5 \\  

$^{180}$Hf & 1.0 & 0.16 & 8.7 & 0.16 & 0.274 & 13 & 44 & 12 & 5 &  9 & 0.05 & 3.31 & 3.30 & 24.3 & 24.4 & 12.9 & 12.9 & 16.7 & 16.6 \\

$^{182}$W  & 1.4 & 0.20 & 8.6 & 0.28 & 0.251 & 31 & 50 & 18 & 6 & 13 & 0.26 & 3.29 & 3.28 & 47.4 & 48.0 & 12.2 & 12.4 & 17.7 & 17.6 \\
$^{184}$W  & 0.9 & 0.18 & 5.4 & 0.16 & 0.236 & 15 & 43 & 10 & 6 &  9 & 0.06 & 3.27 & 3.28 & 16.7 & 16.8 &  8.1 &  8.0 & 13.3 & 13.4 \\
$^{186}$W  & 1.0 & 0.19 & 4.1 & 0.19 & 0.226 & 20 & 43 & 14 & 6 & 11 & 0.09 & 3.23 & 3.25 & 29.1 & 29.2 &  6.0 &  6.1 & 11.4 & 11.4 \\

$^{186}$Os & 1.7 & 0.22 & 4.2 & 0.37 & 0.200 & 72 & 44 & 14 &13 & 18 & 0.20 & 3.17 & 3.19 & 25.9 & 26.1 &  5.6 &  6.1 & 26.5 & 26.8 \\
$^{188}$Os & 1.4 & 0.23 & 3.0 & 0.32 & 0.186 & 57 & 43 & 12 & 7 & 11 & 0.18 & 3.08 & 3.14 & 18.4 & 18.7 &  4.1 &  4.4 & 10.9 & 10.6 \\
$^{190}$Os & 1.4 & 0.24 & 2.2 & 0.34 & 0.178 & 62 & 43 & 10 & 6 &  9 & 0.20 & 2.93 & 3.06 & 12.6 & 12.7 &  3.0 &  3.3 &  7.9 &  7.6 \\ 

$^{238}$Pu & 1.7 & 0.21 &16.5 & 0.36 & 0.286 & 35 & 60 & 26 & 4 & 15 & 0.34 & 3.31 & 3.30 & 96.8 & 97.6 & 23.3 & 23.4 & 25.5 & 25.5 \\
\br
\end{tabular}
\end{table}

\newpage 

\begin{table}

\caption{\label{tab4}Comparison of theoretical predictions of the Bohr Hamiltonian 
with the Woods--Saxon potential to experimental data \cite{ENSDF}
for 9 $\gamma$-unstable Xe isotopes with $R_{4/2}$ $<$ 2.6 
and known $0_2^+$ and $2_{\gamma}^+$ states. 
The angular momenta of the highest levels of the ground state
and quasi-$\gamma_1$ bands included in the rms fit are labelled by $L_g$
and $L_\gamma$ respectively, while $N$ indicates the
total number of levels involved in the fit and $\sigma$ is the
quality measure of (\ref{eq:e99}). All energies are normalized 
to the energy of the first excited state, $E(2^+_1)$. 
For each band, the $R_{4/2}=E(4_1^+)/E(2_1^+)$ ratio (labelled by $4/2$),
the normalized bandhead of the quasi-$\gamma_1$ band (labelled as $2_\gamma/2$), 
and the normalized last members of the ground state and quasi-$\gamma_1$
 bands included in the fit 
(labelled by $L_g/2$ and $L_\gamma/2$ respectively), are reported. 
$\beta_0$ has been obtained from \cite{Raman}.
The theoretical predictions are obtained from  (\ref{Efit1}).
See subsection \ref{numresgunst} for further discussion. }

\tiny
\begin{tabular}{@{} l l l l l l l l l l l l l l l l l l l}
\br
 & & & & & & & & & & & &  & & & & & &  \\

nucleus & $\tilde \beta_0$ & $\tilde A$ & $10^3 \tilde a_0$ & $\beta_0$ & $V_0$ & $10^3 a_0$ & $L_g$ & $L_\gamma$ & $N$ & $\sigma$ &  
$4/2$ & $4/2$ & $L_g/2$ & $L_g/2$ & $2_\gamma/2$ & $2_\gamma/2$ & $L_\gamma/2$ & $L_\gamma/2$ \\
 & & & & & & & & & & & exp & th & exp & th & exp & th & exp & th \\

\mr

$^{118}$Xe & 0.13 & 0.18 &  23 & 0.265 &  0.2 & 48 & 14 & 8 & 13 & 0.26 & 2.40 & 2.34 & 12.9 & 13.1 & 2.8 & 2.3 & 7.8 & 8.0 \\
$^{120}$Xe & 0.21 & 0.17 &  36 & 0.291 &  0.5 & 50 & 14 & 9 & 14 & 0.47 & 2.47 & 2.37 & 13.8 & 13.6 & 2.7 & 2.4 & 9.8 &10.8 \\
$^{122}$Xe & 0.52 & 0.14 &  71 & 0.259 &  4.0 & 35 & 16 & 9 & 15 & 0.65 & 2.50 & 2.32 & 17.5 & 17.6 & 2.5 & 2.3 & 9.7 &11.1 \\
$^{124}$Xe & 0.57 & 0.16 &  88 & 0.212 &  7.2 & 33 & 12 & 8 & 12 & 0.51 & 2.48 & 2.32 & 11.0 & 10.8 & 2.4 & 2.3 & 8.2 & 8.2 \\
$^{126}$Xe & 0.65 & 0.18 & 115 & 0.188 & 11.9 & 33 & 12 & 9 & 13 & 0.63 & 2.42 & 2.28 & 11.0 & 10.1 & 2.3 & 2.3 & 9.1 &10.1 \\
$^{128}$Xe & 0.67 & 0.21 & 143 & 0.184 & 13.4 & 39 & 10 & 7 & 10 & 0.47 & 2.33 & 2.20 &  7.6 &  6.9 & 2.2 & 2.2 & 6.2 & 6.9 \\
$^{130}$Xe & 0.92 & 0.22 & 203 & 0.169 & 29.4 & 37 & 14 & 5 & 10 & 0.30 & 2.25 & 2.11 &  9.5 &  9.5 & 2.1 & 2.1 & 4.1 & 4.7 \\
$^{132}$Xe & 0.90 & 0.24 & 218 & 0.141 & 41.2 & 34 &  6 & 2 &  3 & 0.34 & 2.16 & 2.05 &  3.2 &  3.2 & 1.9 & 2.1 & 1.9 & 2.1 \\
$^{134}$Xe & 1.10 & 0.27 & 302 & 0.119 & 85.6 & 33 &  8 & 5 &  7 & 0.29 & 2.04 & 1.82 &  3.5 &  3.1 & 1.9 & 1.8 & 2.7 & 3.1 \\

\br
\end{tabular}
\end{table}

\end{document}